\documentclass[a4paper]{article}

\usepackage[utf8]{inputenc}
\usepackage[english]{babel}
\usepackage[pdftex]{graphicx}
\usepackage{amssymb,amsmath,textcomp,mathtools}
\usepackage{array}

\newcommand{\comm}[1]{}

\author{Nickolai Muchnoi\thanks{muchnoi@inp.nsk.su}}
\title{FCC-ee polarimeter}

\begin{document}

\pagenumbering{arabic}
\maketitle
\thispagestyle{empty}

\section{Introduction}
Inverse Compton scattering is the classical way to measure beam polarization in lepton machines.
Eligibility of this method at high energy domain has been successfully tested and proved by LEP~\cite{Placidi:1988nj, Knudsen1991} and HERA~\cite{Barber1993} experiments.
Fast measurement of beam polarization allows to apply the resonant depolarization technique for precise beam energy determination~\cite{Skrinski1989, Arnaudon1992}.

\section{Inverse Compton Scattering}
\begin{figure}[h]
\centering
\includegraphics[width=0.6\textwidth]{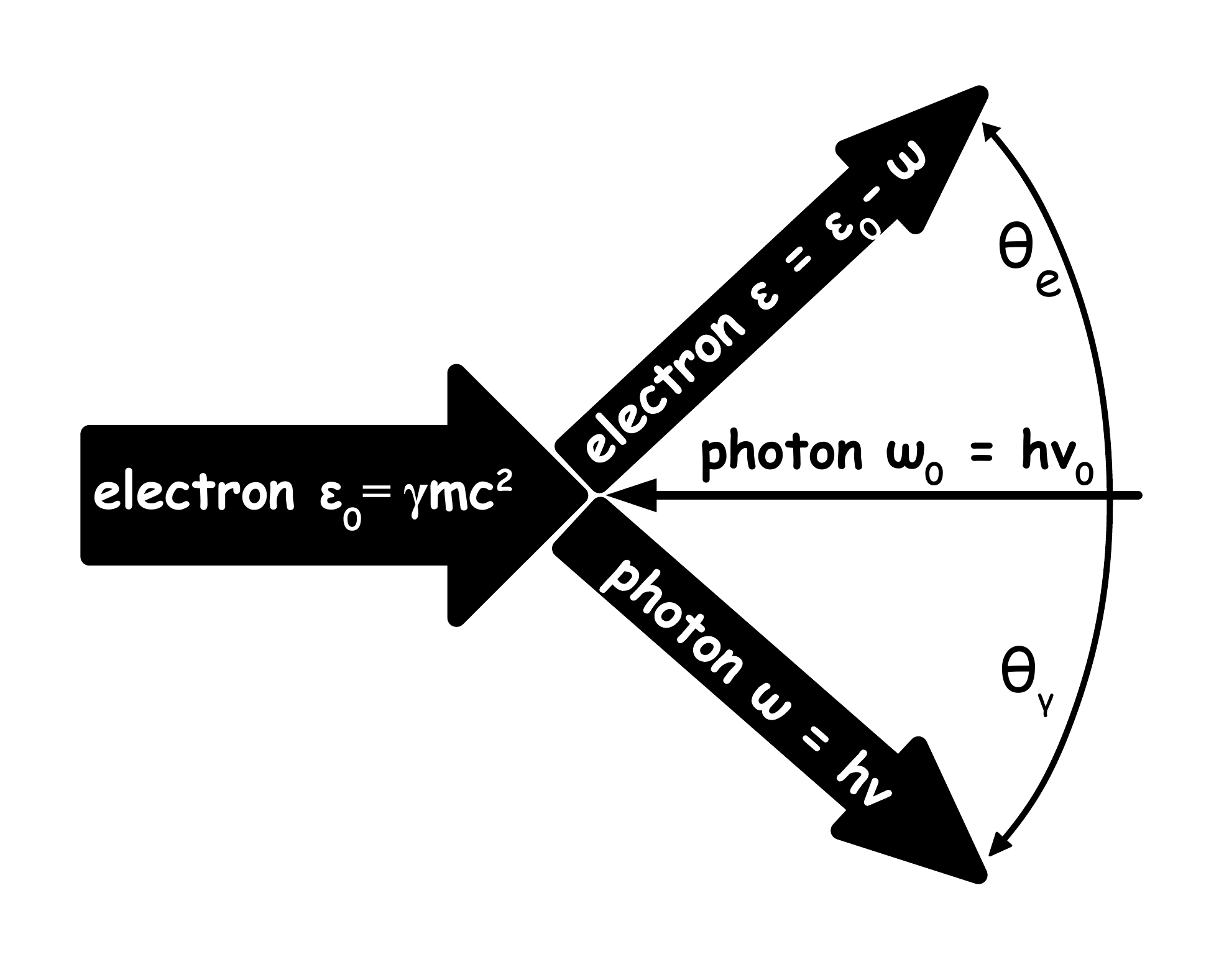}
\caption{Inverse Compton scattering: the thickness of every arrow qualitatively reflects the energy of each particle.
The values of $\omega_0, \varepsilon_0$ and $\omega, \varepsilon$ are the energies of the photon and electron in their initial and final states correspondingly, while $\theta_\gamma$ and $\theta_e$ are the scattering angles of photon and electron.}
\label{fig:kin}
\end{figure}

An illustration for the process of Inverse Compton Scattering (ICS) is presented in Fig.~\ref{fig:kin}.

Considering an ultra-relativistic case ($\varepsilon_0, \varepsilon, \omega \gg \omega_0$) we introduce the universal scattering parameter
\begin{equation}
u
=\displaystyle\frac{\omega}{\varepsilon}
=\displaystyle\frac{\theta_e}{\theta_\gamma}
=\displaystyle\frac{\omega}{\varepsilon_0-\omega}
=\displaystyle\frac{\varepsilon_0-\varepsilon}{\varepsilon},
\label{u}
\end{equation}
bearing in mind the energy and transverse momenta conservation laws while neglecting the corresponding impacts of initial photon.
Parameter $u$ lies within the range $u \in \left[0,\kappa\right]$ and is limited from above by the longitudinal momenta conservation:
 $\kappa$ is twice the initial energy of the photon in the rest frame of the electron, expressed in units of the electron rest energy:
\begin{equation}
\kappa=4\frac{\omega_0\varepsilon_0}{(mc^2)^2} = 2\times 2\gamma\frac{\omega_0}{mc^2}.
\label{kappa}
\end{equation}
If the electron-photon interaction is not head on, the angle of interaction $\alpha\neq\pi$ affects the initial photon energy seen by the electron, and $\kappa$ parameter becomes\footnote{this is correct when $\tan(\alpha/2) \gg 1/\gamma$.}
\begin{equation}
\kappa(\alpha)=4\frac{\omega_0\varepsilon_0}{(mc^2)^2} \sin^2\left(\frac{\alpha}{2}\right).
\label{kappa_any}
\end{equation}

For the FCC-ee polarimeter we consider the interaction of laser radiation with the electrons in the electron beam energy range $\varepsilon_0 \in [45:185]$~GeV.
The energy of the laser photon $\omega_0$ is coupled with the radiation wavelength in vacuum $\lambda_0$: $\omega_0 = hc/\lambda_0$, where $hc = 1.23984193$~eV$\cdot\mu$m.
In particular case when $\lambda_0=1\;\mu$m, $\varepsilon_0=100$~GeV and $\alpha=\pi$ one obtains the ``typical'' value of $\kappa$ parameter for the FCC-ee case, $\kappa \simeq 1.9$.
Maximum energy of backscattered photon $\omega_{max}$ obviously corresponds to the minimal energy of scattered electron $\varepsilon_{min}$, both values are easily obtained from definitions (\ref{u}) -- (\ref{kappa_any}) when $u=\kappa$:
\begin{equation}
\omega_{max} = \frac{\varepsilon_0\kappa}{1+\kappa} \;\text{ and }\; \varepsilon_{min} = \frac{\varepsilon_0}{1+\kappa}.
\label{wmax_emin}
\end{equation}
Note that $\omega_{max}=\varepsilon_{min}$ when $\kappa=1$.
It's not hard to show that the scattering angles of photon $\theta_\gamma$ and electron $\theta_e$ (see Fig.~\ref{fig:kin}) depend on $u$ and $\kappa$ as:
\begin{equation}
\theta_\gamma = \frac{1}{\gamma}\sqrt{\frac{\kappa}{u}-1} \;\text{ and }\; \theta_e = \frac{u}{\gamma}\sqrt{\frac{\kappa}{u}-1}.
\label{thetas}
\end{equation}
The electron scattering angle $\theta_e$ can never exceed the limit $\max(\theta_e)=\kappa/2\cdot\gamma = 2\omega_0/mc^2$ and we see that this value does not depend on $\varepsilon_0$.
Almost any experimental application of the backscattering of laser radiation on the electron beam for any reason implies the use of the following minimal scheme:
\begin{figure}[h]
\centering
\includegraphics[width=0.75\textwidth]{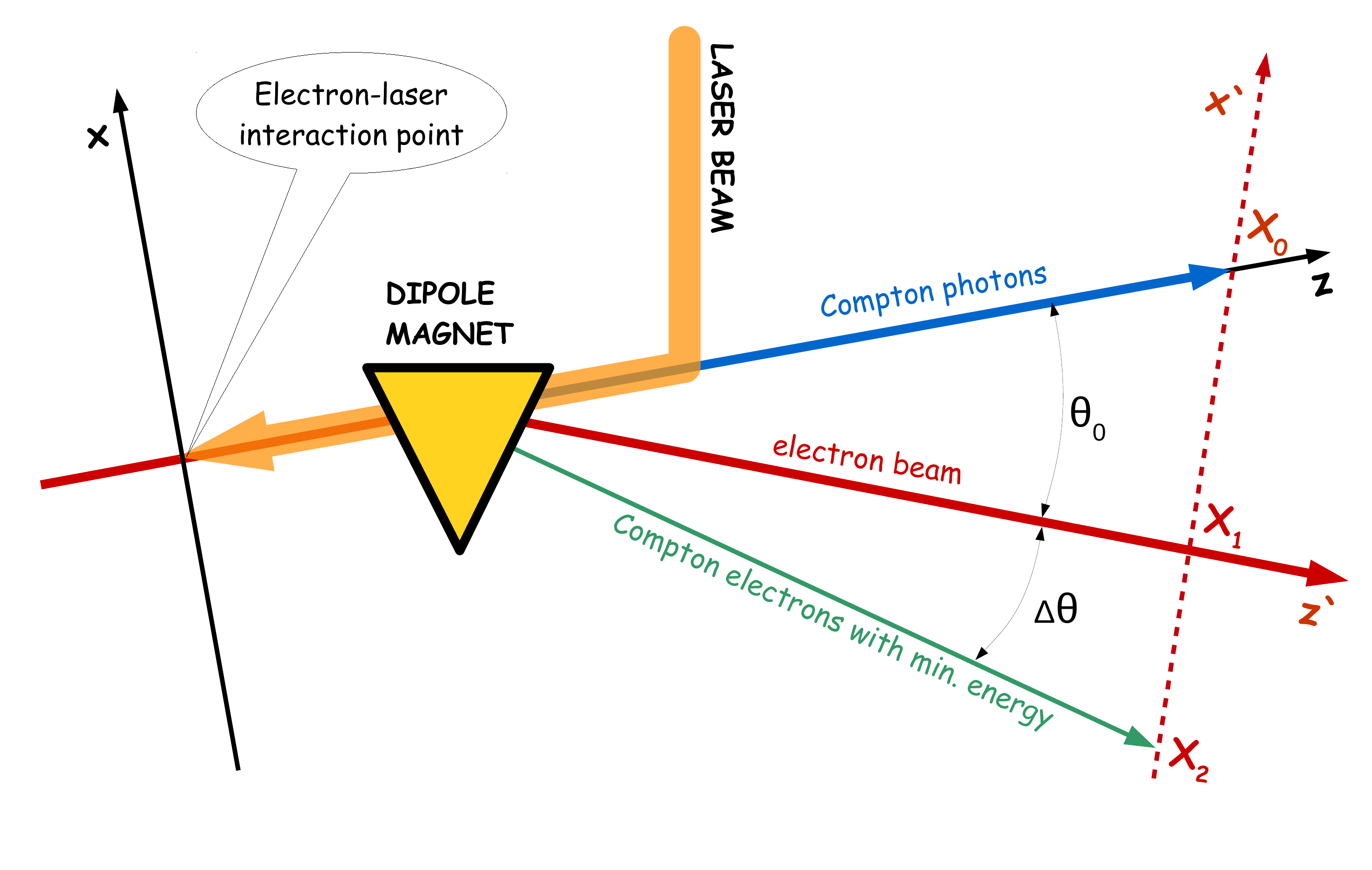}
\caption{Regular layout of ICS experiments realization.}
\label{fig:cxema}
\end{figure}

In Fig.~\ref{fig:cxema} the laser radiation is focused, inserted into the machine vacuum chamber and directed to the interaction point where scattering occurs.
The dipole is used to separate scattered photons (and electrons) from the electron beam, propagating in the machine's vacuum chamber.
$x$-axis and $z$-axis define the coordinate system in the interaction point, the plane of the figure is the plane of machine, the vertical $y$-axis is perpendicular to the plane of figure.
After the dipole, the coordinate system $(x',z')$ is rotated by the beam bending angle $\theta_0$.

\subsection{ICS cross section}

The ICS cross section in general is sensitive to polarization states of all initial and final particles~\cite{berestetskii1982quantum}.
It is common to average the polarization terms of the final states, then the cross section depends solely from the initial photon and electron polarizations.
In order do describe polarization states of the laser and electron beams in the coordinate system $x,y,z$, presented in Fig.~\ref{fig:cxema}, let's introduce modified Stokes parameters.
\begin{itemize}
\label{poldefs}
\item $\xi_\perp \in [0:1]$ and $\varphi_\perp \in [0:\pi]$ are the degree of laser linear polarization and its azimuthal angle.
\item $\xi_\circlearrowright \in [-1:1]$ is the sign and degree of circular polarization of laser radiation: $\sqrt{\xi_\perp^2 + \xi_\circlearrowright^2}=1$.
\item $\zeta_\perp \in [0:1]$ and $\phi_\perp \in [0:2\pi]$ are the degree of transverse e$\pm$ beam polarization and its azimuthal angle.
\item $\zeta_\circlearrowright \in [-1:1]$ is the sign and degree of longitudinal spin polarization of the electrons: $\sqrt{\zeta_\perp^2 + \zeta_\circlearrowright^2} \in [0:1]$.
\end{itemize}
Then, the ICS cross section is described by the sum of three terms: $d\sigma = d\sigma_0 + d\sigma_\parallel + d\sigma_\perp$, these terms are: $d\sigma_0$ -- unpolarized electron; $d\sigma_\parallel$ -- longitudinal electron polarization; $d\sigma_\perp$ -- transverse electron polarization:
\begin{equation}
\begin{aligned}
d\sigma_0  & = & \frac{r_e^2}{\kappa^2(1+u)^3} &
\left( \kappa (1 + (1+u)^2) - 4\frac{u}{\kappa}(1+u)(\kappa-u)\Bigl[ 1-\xi_\perp \cos\bigl(2(\varphi-\varphi_\perp)\bigr) \Bigr] \right)   & du\;d\varphi,\\
d\sigma_\parallel & = & \frac{\xi_\circlearrowright \zeta_\circlearrowright r_e^2}{\kappa^2(1+u)^3} &
\;\;\;\;\;\;\;\; u(u+2)(\kappa-2u) & du\;d\varphi ,\\
d\sigma_\perp & = & - \frac{ \xi_\circlearrowright \zeta_\perp r_e^2}{\kappa^2(1+u)^3} &
\;\;\;\;\;\;\;\; 2u\sqrt{u(\kappa-u)} \cos(\varphi-\phi_\perp) & du\;d\varphi .
\end{aligned}
\label{suall}
\end{equation}
In equations~(\ref{suall}) $r_e$ is the classical electron radius and $\varphi$ is the observer's azimuthal angle.
As one can see from equations~(\ref{suall}), the last term $d\sigma_\perp$, most important for FCC-ee polarimeter, can not modify the total cross section, which in absence of longitudinal polarization of electrons is obtained by integration of $d\sigma_0$ only:
\begin{equation}
\sigma_0( \kappa)  =
\frac{2 \pi r_e^2}{\kappa} \left( \left[ 1-\frac{4}{\kappa}-\frac{8}{\kappa^2}\right] \mathrm{log}(1+\kappa)
+ \frac{1}{2}\left[ 1-\frac{1}{( 1+\kappa)^2}\right] +\frac{8}{\kappa}\right).
\label{tcs}
\end{equation}
In case when $\kappa \ll 1$ expression (\ref{tcs}) tends to the Thomson cross section $\sigma_0=\frac{8}{3}\pi r_e^2 \left( 1-\kappa\right)$.

The above expressions are enough e.~g. to start Monte-Carlo generator and allow further analysis of scattered particles distributions.
Dimensionless parameter $u\in[0:\kappa]$ is obtained according to the initial values of  $\varepsilon_0$, $\omega_0$, $\alpha$ and polarization koefficients. The probability distribution of $u$ is defined by the cross section (\ref{suall}).
Then the required properties, like $\omega$, $\varepsilon$, $\theta_e$ or $\theta_\gamma$ are obtained using equations (\ref{u}) and (\ref{thetas}).
However, the influence of bending magnet in Fig.~\ref{fig:cxema} on scattered electrons is not yet considered.

\subsection{Bending of electrons}

Let's describe the dipole strength by the parameter $B$, assuming for the sake of brevity that it is proportional to the integral of magnetic field along the electron trajectory.
The electron with energy $E$ will be bent to the angle $\theta=B/E$ under the assumption that $B$ is the same for all energies under consideration
\footnote{The validity of this assumption will be discussed on page \pageref{DeltaL}.}.
By equation (\ref{u}) we express the energy $\varepsilon$ of scattered electron through the ICS parameter $u$: $\varepsilon=\varepsilon_0/(1+u)$.
This electron is bent by the dipole to the angle
\begin{equation}
\theta = \frac{B}{\varepsilon} =
\frac{B}{\varepsilon_0} + u\frac{B}{\varepsilon_0}=
\theta_0 + u\theta_0,
\label{bend}
\end{equation}
i.~e. $\theta$ is the sum of the beam bending angle $\theta_0$ and the bending angle $\Delta\theta = u\theta_0$, caused by electron energy loss in ICS.
Both $\theta_0$ and $\Delta\theta$ are shown in Fig.~\ref{fig:cxema} for the maximum possible $u$ value $u=\kappa$.
Note that $\kappa\theta_0$ does not depend on $\varepsilon_0$.
In ref.~\cite{Muchnoi2009} it was suggested to use the ratio $\Delta\theta/\theta_0=\kappa$ for the ILC beam energy determination.

Let us introduce a new designation $\vartheta_x \equiv \gamma(\theta-\theta_0) = u\vartheta_0$ which is the angle $\Delta\theta$, measured in units of $1/\gamma$.
The scattering angle of an electron due to ICS, expressed in the same units, is $\vartheta = u\sqrt{\kappa/u-1}$ as follows from eq.~(\ref{thetas}).
By splitting $\vartheta$ into $x$ and $y$ components and gathering all angles together we get:
\begin{equation}
\begin{aligned}
\vartheta_x  = & \sqrt{u(\kappa-u)}\cos\varphi + u\vartheta_0, \\
\vartheta_y  = & \sqrt{u(\kappa-u)}\sin\varphi.
\end{aligned}
\label{txty}
\end{equation}

Since the backscattered photons are not bent by the dipole, the photon transverse angles (eq.~(\ref{thetas})) in the same space and in the same units as in eq.~(\ref{txty}), according to similar considerations, are the following:
\begin{equation}
\begin{aligned}
\eta_x  = & - \sqrt{\kappa/u-1}\cos\varphi - \vartheta_0, \\
\eta_y  = & - \sqrt{\kappa/u-1}\sin\varphi.
\end{aligned}
\label{nxny}
\end{equation}

\section{Polarimeter location and layout}

\begin{figure}[h]
\centering
\includegraphics[width=0.75\textwidth, height=6cm]{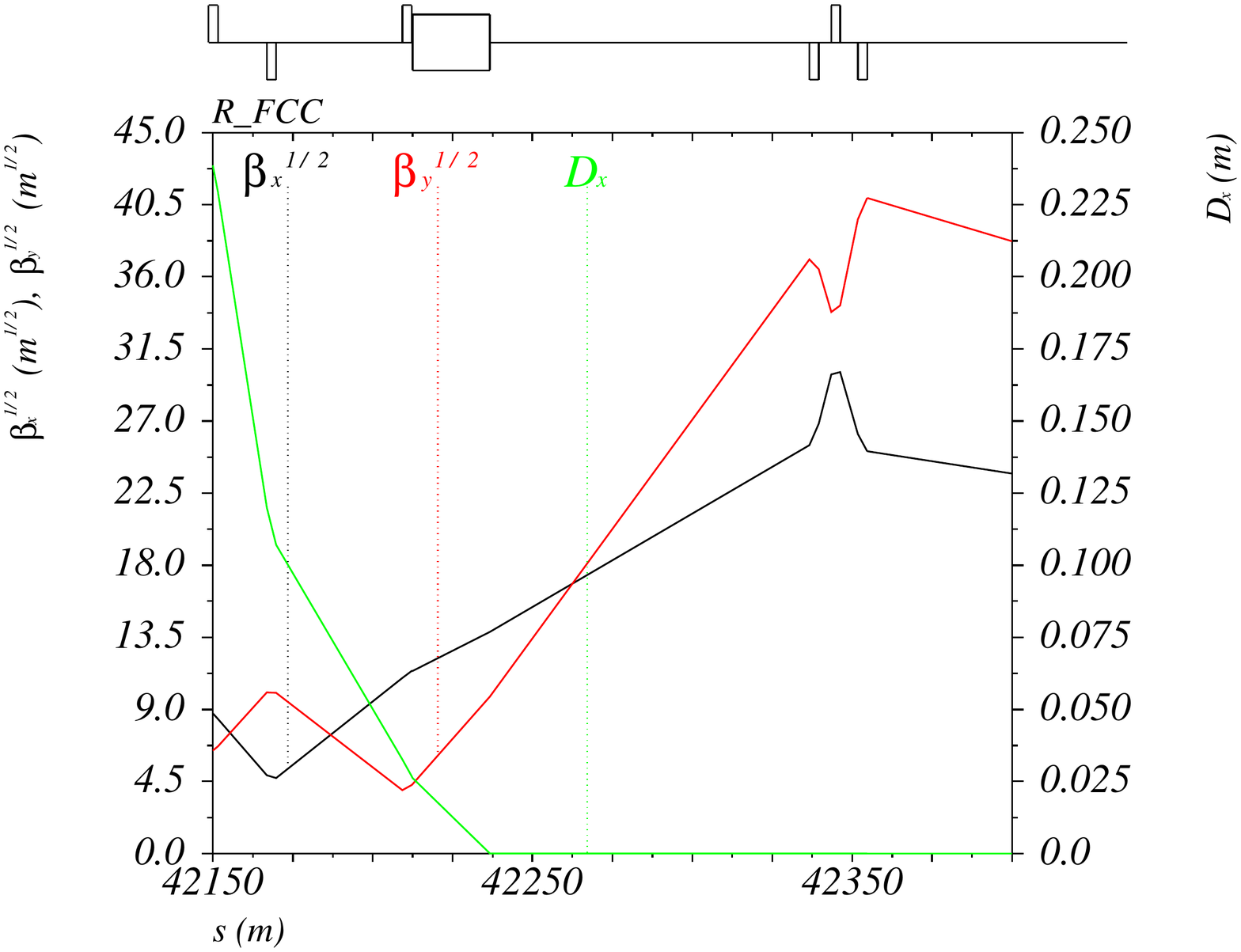}
\caption{Polarimeter location with respect to FCC-ee lattice.}
\label{fig:madx}
\end{figure}

\begin{figure}[h]
\centering
\includegraphics[width=0.75\textwidth]{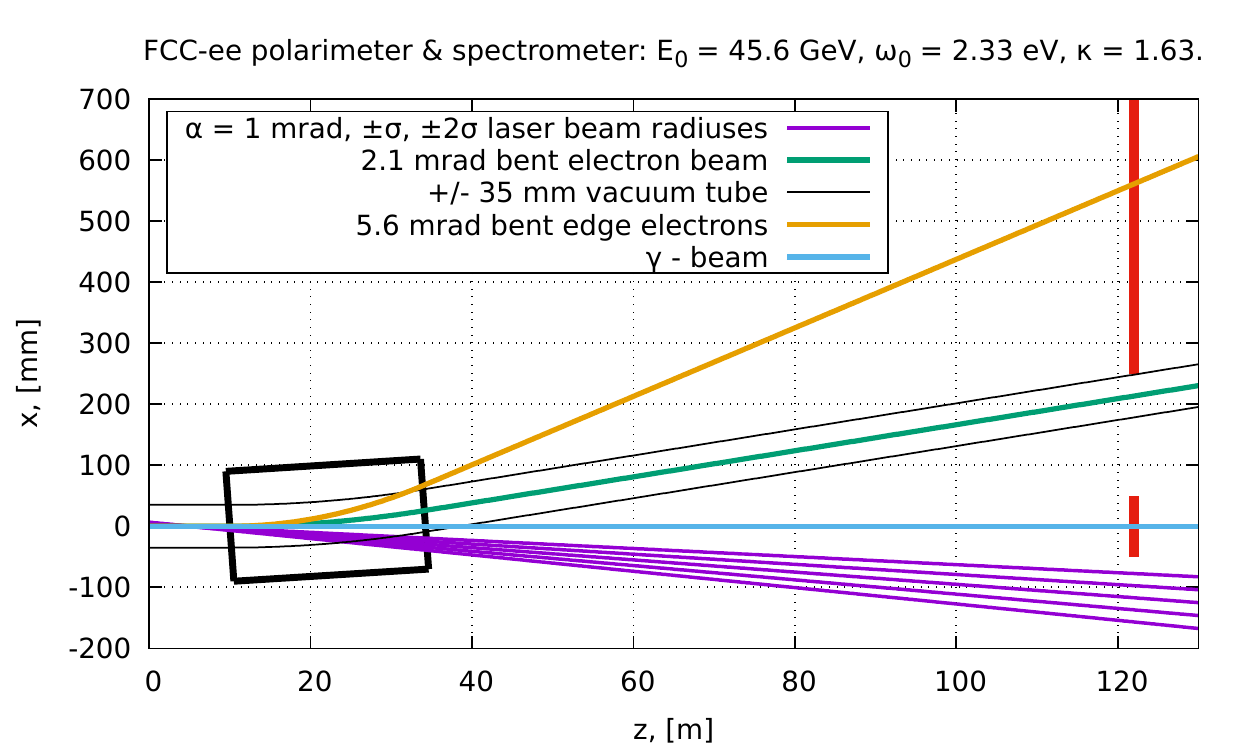}
\caption{Sketch of the polarimeter: dipole ($L=24.12$~m, $\theta_0=2.13$~mrad, $B=0.0135$~T, $R_0=11302$~m), vacuum chamber, particle trajectories.
Red vertical bars on the right side show the location of the scattered particles detectors -- 100~m away from the center of the dipole.}
\label{fig:100m}
\end{figure}

The polarimeter will be installed in the FCC-ee section shown in Fig.~\ref{fig:madx}.
After the dispersion suppressing dipole magnet, about 100~m of free beam propagation is reserved to allow clear separation of the ICS photons and electrons from the beam.
Red bars from the right side are the detectors of scattered electrons and photons, which
The interaction of the pulsed laser beam with the electron beam occurs just between the dipole and preceding quadrupole, where there is a local minimum of vertical $\beta$-function.
In Fig.~\ref{fig:100m} there is the sketch of the polarimeter apparatus arrangement in horizontal plane.
The laser radiation $\lambda_0=532$~nm is inserted to the vacuum chamber from the right side and focused to the interaction point ($z=5$~m), the laser spot transverse size at i.~p. is $\sigma_0=0.25$~mm.
According to Fig.~\ref{fig:100m}, laser-electron interaction angle is $\alpha = \pi - 0.001$ and the relative difference between $\kappa$ from eq.~(\ref{kappa}) and $\kappa(\alpha)$ from eq.~(\ref{kappa_any}) is as small as $2.5\cdot10^{-7}$.

\subsection{Spectrometer}
Figure~\ref{fig:100m} helps to understand how much could be the difference of the B-field integral, seen by the electrons with different energies.
All of the electrons enter the dipole of length $L$ along the same line -- the beam orbit.
Then, the radius of trajectory will be dependent on the electron energy.
Let $R_0$ to be the radius of an electron with energy $\varepsilon_0$ and $\theta_0 = L/R_0$ is the beam bending angle.
The minimal radius of an electron after scattering on the laser light will be $R_0/(1+\kappa)$.
After passing the dipole these two electrons will have the difference $\Delta x \simeq \kappa L \theta_0 /2 $ in transverse horizontal coordinates.
With the parameters of Fig.~\ref{fig:100m} this difference is $\Delta x \simeq 43$~mm.
The length of the trajectories of these two electrons inside the dipole will be also different, i.~e. even in case of absolutely uniform dipole their field integrals will not be the same.
In case of rectangular dipole, the exact expression for relative difference of the lengths of trajectories is:
\begin{equation}
\frac{\Delta L}{L} = \frac{2}{\theta_0}
\left[
\frac{1}{1+\kappa} \arcsin \left( \frac{\theta_0}{2} (1+\kappa) \sqrt{1+ \left(\frac{\kappa\theta_0}{2}\right)^2} \right)
- \arcsin \left( \frac{\theta_0}{2} \right) \right].
\label{DeltaL}
\end{equation}
As we see this relative difference depends on $\theta_0$ and $\kappa$ only.
With the set of parameters taken from Fig.~\ref{fig:100m}, i.~e. $\theta_0=2.13$~mrad and $\kappa=1.63$, $\Delta L/L = 2.63\cdot10^{-6}$.

The result of this section is the proof of the validity of assumption about the equality of the integrals of the magnetic field for the electron beam and scattered electrons.
This assumption was found to be rather accurate for the dipole with perfectly uniform field, however {\em shorter dipole is much more preferable} in order to decrease $\Delta x$ and hence have less concerns about the field quality.

\subsection{Monte Carlo distributions of scattered particles}
\begin{figure}[h]
\centering
\includegraphics[width=\textwidth]{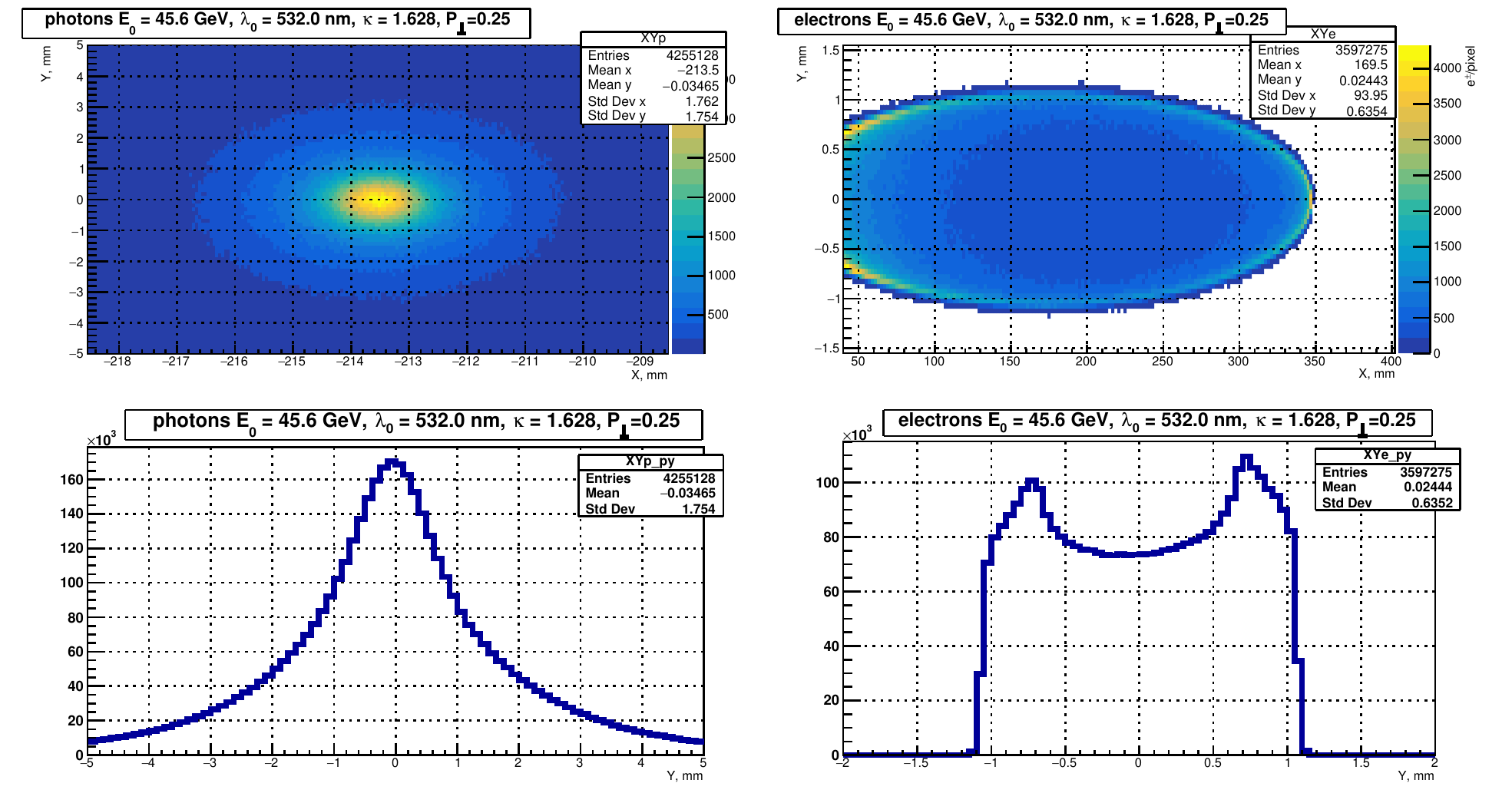}
\caption{MC results for $P_\perp = \xi_\circlearrowright \zeta_\perp = 0.25$ and $\phi_\perp=\pi/2$.}
\label{fig:pos25}
\end{figure}

\begin{figure}[h!]
\centering
\includegraphics[width=\textwidth]{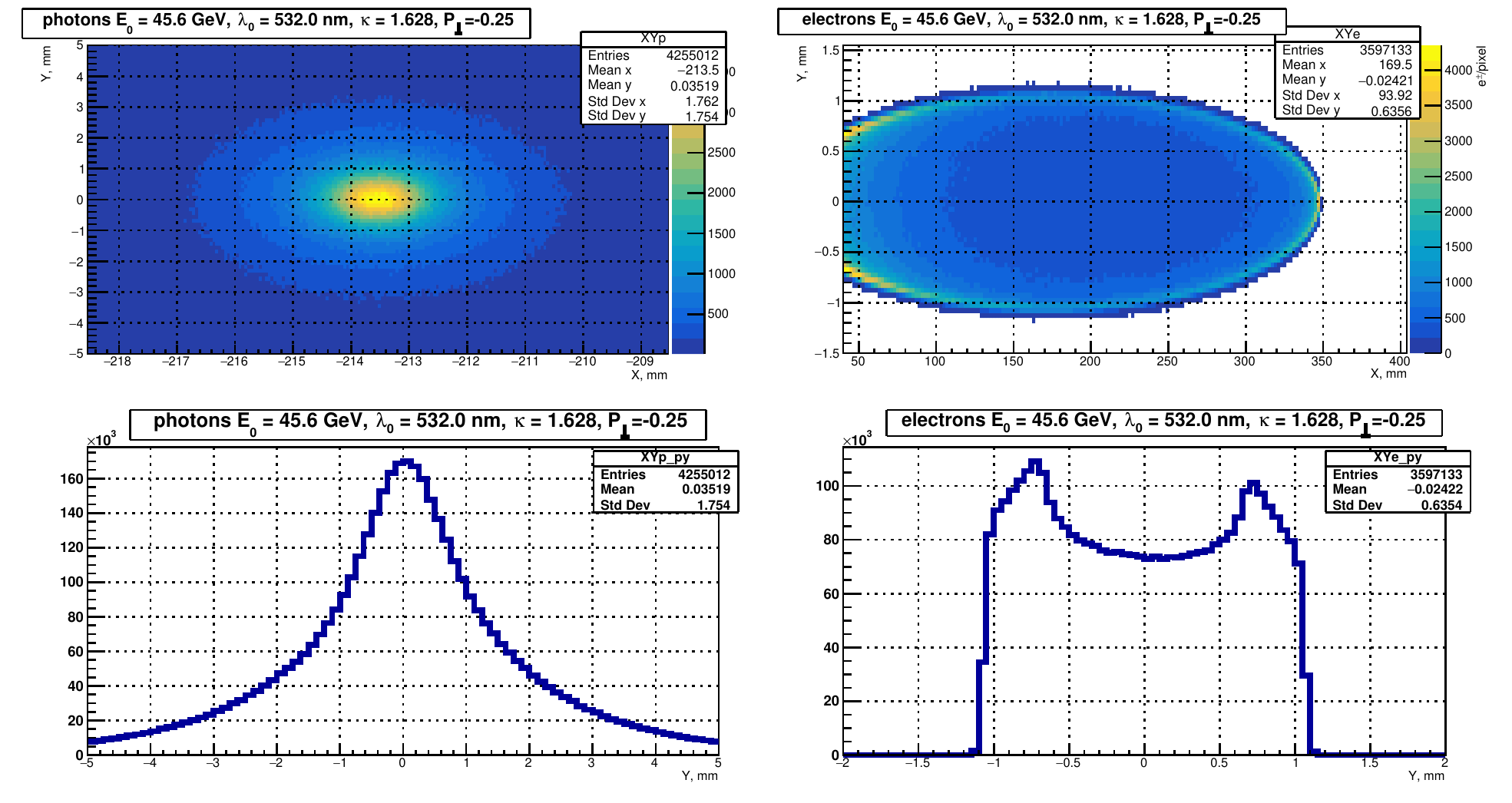}
\caption{MC results for $P_\perp = \xi_\circlearrowright \zeta_\perp = -0.25$ and $\phi_\perp=\pi/2$.}
\label{fig:neg25}
\end{figure}

\begin{figure}[t]
\centering
\includegraphics[width=\textwidth]{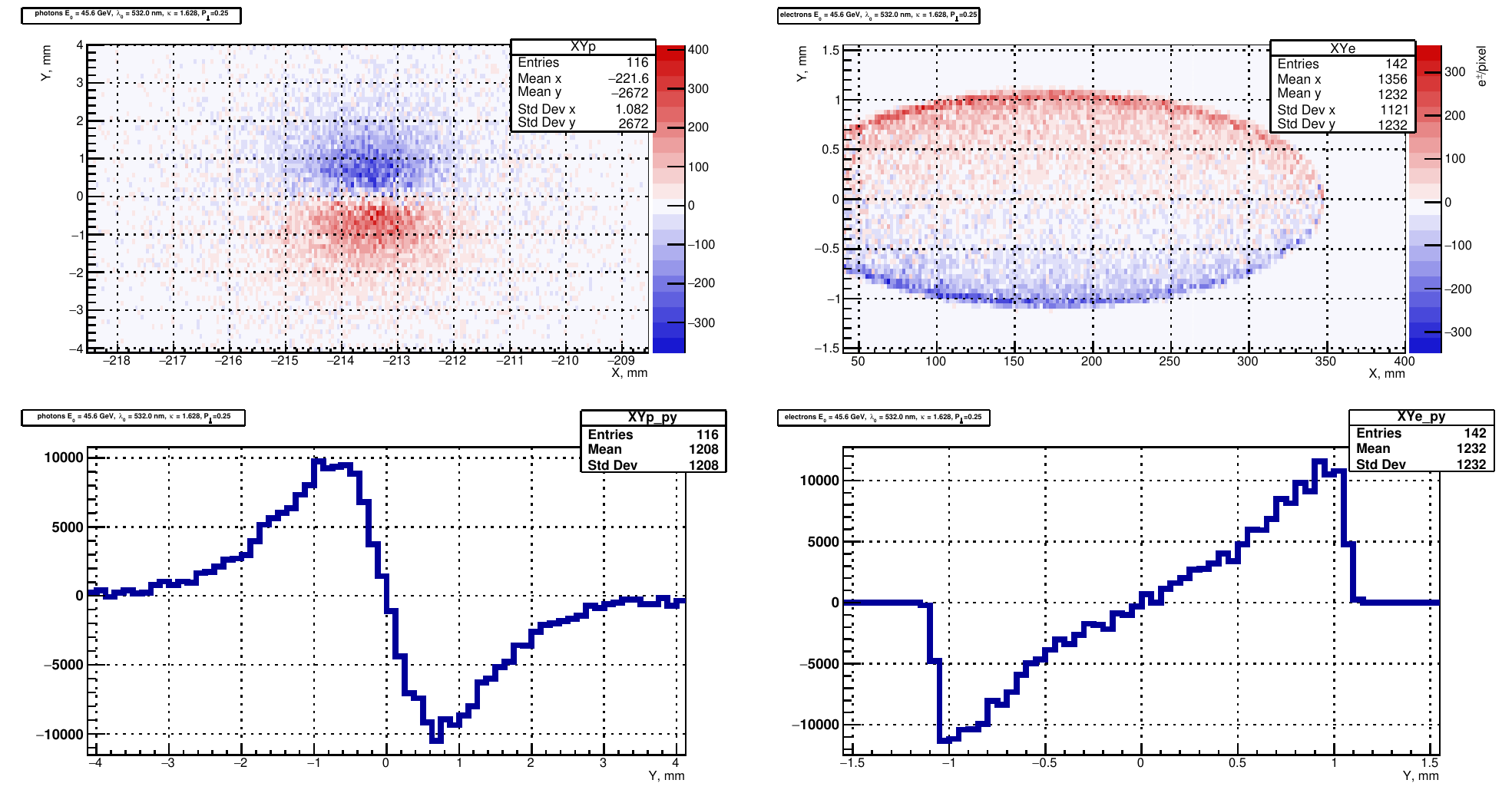}
\caption{The difference between corresponding distributions in Fig.~\ref{fig:pos25} and Fig.~\ref{fig:neg25}.}
\label{fig:thedifference}
\end{figure}

The MC generator was created to obtain the 2D ($x,y$) distributions of scattered photons and electrons at the detectors, located as it is shown in Fig.~\ref{fig:100m}.
The ICS parameters are: $\varepsilon_0=45.6$~GeV and $\omega_0=2.33$~eV.
The spectrometer configuration is described by the beam bending angle $\theta_0=2.134$~mrad, the lengths of the dipole $L=24.12$~m and two spectrometer arms.
First arm $L_1=117$~m  is the distance between laser-electron IP and the detector.
Second arm  $L_2=100$~m  is the distance between the longitudinal center of the dipole and the detector.
The impact of the electron beam parameters is accounted by introducing the angular spreads according to the beam emittances $\epsilon_x=0.27$~nm and $\epsilon_y=1$~pm.
The horizontal and vertical electron angles $x'$ and $y'$ in the beam are described by normal distributions with means equal to zero and standard deviations $\sigma_x = \sqrt{\epsilon_x/\beta_x}$ and $\sigma_y = \sqrt{\epsilon_y/\beta_y}$.
The values of $\beta$-functions were taken from Fig.~\ref{fig:madx}.
The procedure is:
\begin{itemize}
\item raffle $u\in[0,\kappa]$ and $\varphi\in[0:2\pi]$ according to 2D function $d\sigma(u,\varphi)$ (eq.~(\ref{suall})),
\item raffle $x'$ and $y'$ according to corresponding normal distributions,
\item obtain photon $X_\gamma, Y_\gamma$ and electron $X_e, Y_e$ transverse coordinates at the detection plane:
\end{itemize}
\begin{equation}
\begin{aligned}
X_\gamma & =  x' L_1 - \frac{L_1}{\gamma}\sqrt{ \kappa/u-1} \cos \varphi - \theta_0 L_2, \\
Y_\gamma & =  y' L_1 - \frac{L_1}{\gamma}\sqrt{ \kappa/u-1} \sin \varphi, \\
X_e      & =  x' L_1 + \frac{L_1}{\gamma}\sqrt{u(\kappa-u)} \cos \varphi + u \theta_0 L_2, \\
Y_e      & =  y' L_1 + \frac{L_1}{\gamma}\sqrt{u(\kappa-u)} \sin \varphi.
\end{aligned}
\end{equation}
The results of such a simulation for an electron beam with $\zeta_\perp=25$\% vertical ($\phi_\perp=\pi/2$) spin polarization are presented in Fig.~\ref{fig:pos25} and Fig.~\ref{fig:neg25}.
The difference between the figures is the laser polarization $\xi_\circlearrowright=+1$ (Fig.~\ref{fig:pos25}) and  $\xi_\circlearrowright=-1$ (Fig.~\ref{fig:neg25}).
The 2D distributions for both photons and electrons are plotted along the same horizontal axis $x$, where $x=0$ corresponds to the position of the electron beam.
The detectors for scattered particles are located outside the machine vacuum chamber.
The scattered electrons distribution starts form $x = 40$~mm: this is the radius of the vacuum chamber.
The 1D distributions in the bottom of each figure are the projections of 2D distributions to the vertical axis $y$.
The mean $y$-values of these distributions are shifted up or down from zero according to the presence of beam polarization and corresponding asymmetries in ICS cross section.
In Fig.~\ref{fig:thedifference} all distributions are obtained by subtraction of corresponding distributions from Fig.~\ref{fig:pos25} and Fig.~\ref{fig:neg25}.
Detecting the up-down asymmetry in the distribution of laser backscattered photons is a classical way to measure the transverse polarization of the electron beam.
In \cite{Mordechai:2013zwm} it was proposed to use the up-down asymmetry in the distribution of scattered electrons for the transverse polarization measurement at the ILC.
It was suggested to measure the distribution of scattered electrons by Silicon pixel detector.

Maximum up-down asymmetry in the distribution of scattered electrons occurs at the scattering angles of $\theta^*_e = \pm 2\omega_0/mc^2$, which is approximately $\pm 9\;\mu$rad (see eqs.~(\ref{thetas}) and (\ref{suall})).
Asymmetry can be observed only if the distribution is not blurred by the electron beam emittance.
On the other hand, maximum up-down asymmetry in the distribution of scattered photons occurs at the scattering angles of $\theta^*_\gamma \simeq 1/\gamma$ which is almost the same as $\theta^*_e$ in our particular case.
But e.~g. when beam energy is about 5~GeV, $\theta^*_\gamma$ is ten times larger then $\theta^*_e$ and the measurement of beam polarization by photons looks like more preferable.
What are the benefits of scattered electrons against scattered photons for the FCC-ee polarimeter?
\begin{itemize}
\item Scattered electrons propagate to the inner side of the beam orbit, i.~e. there is no direct background from high energy synchrotron radiation.
\item Unlike photons, charged electrons are ready to be detected by their ionization losses. The photons need to be converted to e$^+$e$^-$ pairs: this leads either to low detection efficiency either to low spatial resolution.
\item Despite the fact that the fluxes of scattered photons and electrons are the same, the flux density of electrons is much lower due to bending and corresponding spatial separation by energies. Simultaneous detection of multiple scattered electrons thus is much easier.
\item Analysis of the scattered electrons distribution allows to measure the longitudinal beam polarization as well as the transverse one.
\item As one can observe from Figs.~\ref{fig:pos25}--\ref{fig:thedifference}, change of laser circular polarization leads to a redistribution of the scattered electron density within a fixed elliptic shape of distribution. This fact potentially provides better systematic accuracy for beam polarization determination.
\end{itemize}
Nevertheless both photon and electron distributions are going be measured by FCC-ee polarimeter.
First, to exploit directly the LEP and HERA experience.
Second, to be able to measure the center of the photon distribution in both $x$ and $y$ dimensions.
The latter is required for direct beam energy determination, which will be discussed below.

\section{The shape of the scattered electrons distribution}

This section owes its origin to the successful application of the method of direct electron beam energy determination by backscattering of laser radiation.
The approach is based on the measurement of $\omega_{max}$ (see eq.~\ref{wmax_emin}) in cases when this energy can be measured with good accuracy and in absolute scale.
For the last years, the positive experience on application of this method is accumulated at the low energy colliders VEPP-4M, BEPC-II and VEPP-2000~\cite{Achasov:2017abc}.
Despite the fact that this method is not directly applicable in FCC-ee case, let us try to figure out what can be learned from the elliptical shape of the distribution of scattered electrons, obtained my MC simulations above.

We return to the consideration of the spatial distribution of the scattered electrons.
From (\ref{txty}) we obtain the square equation on $u$:
\begin{equation}
(\vartheta_x-u\vartheta_0)^2 + \vartheta_y^2 = u(\kappa-u),
\label{nequality}
\end{equation}
with the roots equal to:
\begin{equation}
u^\pm = \frac{\kappa+2\vartheta_0\vartheta_x\pm\sqrt{\kappa^2-4(\vartheta_x^2+\vartheta_y^2(1+\vartheta_0^2)-\kappa\vartheta_0\vartheta_x)}}{2(1+\vartheta_0^2)}.
\label{upm}
\end{equation}
The average value of $u$ and its limiting value for the large values of $\vartheta_0$ do not depend on $\vartheta_y$:
\begin{equation}
\langle u \rangle
= \frac{u^+ + u^-}{2}
= \frac{\kappa/2+\vartheta_0\vartheta_x}{1+\vartheta_0^2} \;\;\xrightarrow{\vartheta_0\gg1}\;\; \frac{\vartheta_x}{\vartheta_0}.
\label{nutxty-one}
\end{equation}
In the $\vartheta_x,\vartheta_y$ plane all the scattered electrons are located inside the ellipse (what we have seen in Figs.~\ref{fig:pos25}, \ref{fig:neg25}), described by the radicand in eq.~(\ref{upm}).
The center of the ellipse is located at the point $[\vartheta_x =\kappa\vartheta_0/2; \vartheta_y=0]$, its horizontal semiaxis $A=\kappa\sqrt{1+\vartheta_0^2}/2$ while the verical (along $\vartheta_y$) is equal $B=\kappa/2$.

In particular, this means that
\begin{equation}
\vartheta_x^{max} = \frac{\kappa}{2}\Bigl(\vartheta_0+\sqrt{1+\vartheta_0^2}\Bigr)\xrightarrow{\vartheta_0\gg1}\kappa\vartheta_0.
\label{varthetamax}
\end{equation}
Recall that according to notation introduced above, $\vartheta$-s are the angles measured in units of $1/\gamma$, while $\theta$-s are the angles in radians.
In radians expression (\ref{varthetamax}) looks like
\begin{equation}
\Delta\theta = \frac{\kappa}{2}\Bigl(\theta_0 + \sqrt{1/\gamma^2+\theta_0^2}\Bigr)\xrightarrow{\theta_0\gg1/\gamma}\kappa\theta_0,
\label{thetamax}
\end{equation}
where $\Delta\theta$ and $\theta_0$ were presented in Fig.~\ref{fig:cxema}.
In order to transform the ICS cross section from $u,\varphi$ variables to $\vartheta_x,\vartheta_y$ we calculate the Jacobian matrix:
\begin{equation}
\mathbf{J}=
\begin{bmatrix}
\displaystyle\frac{\partial\vartheta_x}{\partial u} & \displaystyle\frac{\partial\vartheta_x}{\partial\varphi}\\[1em]
\displaystyle\frac{\partial\vartheta_y}{\partial u} & \displaystyle\frac{\partial\vartheta_y}{\partial\varphi}
\end{bmatrix}
=
\begin{bmatrix*}[r]
\displaystyle \vartheta_0 + \frac{\kappa/2-u}{\sqrt{u(\kappa-u)}}\cos\varphi &
\displaystyle -\sqrt{u(\kappa-u)}\sin\varphi \\[1em]
\displaystyle  \frac{\kappa/2-u}{\sqrt{u(\kappa-u)}}\sin\varphi &
\displaystyle  \sqrt{u(\kappa-u)}\cos\varphi
\end{bmatrix*}.
\label{nJacob}
\end{equation}
The matrix determinant is:
\begin{equation}
\det(\mathbf{J}) = \kappa/2 - u + \vartheta_0\sqrt{u(\kappa-u)}\cos\varphi = \sqrt{\kappa^2/4-\vartheta_x^2-\vartheta_y^2(1+\vartheta_0^2)+\kappa\vartheta_0\vartheta_x} .
\end{equation}
Hence, $du d\varphi = 2 d\vartheta_xd\vartheta_y/\det(\mathbf{J})$, where ``2'' is due to the sum of ``up'' and ``down'' solutions of eq.~(\ref{upm}).
Let us perform another change of variables: instead of $\vartheta_x,\vartheta_y$ we introduce $x$ and $y$.
With this new variables the cross section exists inside the circle of radius $R=1$ centered at $(x=0;y=0)$:
\begin{equation}
x = \frac{\vartheta_x-\kappa\vartheta_0/2}{\kappa/2\sqrt{1+\vartheta_0^2}},\;\;\;\;\;
y = \frac{\vartheta_y}{\kappa/2}.
\label{nxy}
\end{equation}
Then:
\begin{equation}
\begin{aligned}
du d\varphi & = & \frac{\kappa\,dx\,dy}{\sqrt{1 - x^2 - y^2}} ,\\
u = \langle u \rangle  &=& \frac{\kappa}{2} \left( 1 + \frac{x\vartheta_0}{\sqrt{1+\vartheta_0^2}}\right),\\
\sin(\varphi) & = & \frac{y\, \kappa}{2\sqrt{u(\kappa-u)}}.
\end{aligned}
\label{janduxy}
\end{equation}
In (\ref{janduxy}) the vertical transverse electron polarization ($\phi_\perp=\pi/2$) is assumed, then $\cos(\varphi-\phi_\perp)=\sin(\varphi)$.
Considering backscattering of circularly polarized laser radiation ($\xi_\circlearrowright = \pm 1$) on the electron beam, where both vertical transverse ($\zeta_\perp \neq 0, \phi_\perp=\pi/2$) and longitudinal ($\zeta_\circlearrowright \neq 0$) polarizations are possible, we rewrite the cross sections (\ref{suall}) in new variables:
\begin{equation}
\begin{aligned}
d\sigma_0  & = & \frac{r_e^2}{\kappa(1+u)^3\sqrt{1 - x^2 - y^2}} &
\left( 1 + (1+u)^2 - 4\frac{u}{\kappa}(1+u)(1-\frac{u}{\kappa}) \right) & dx\;dy,\\
d\sigma_\parallel & = & \frac{\xi_\circlearrowright \zeta_\circlearrowright  r_e^2}{\kappa(1+u)^3\sqrt{1 - x^2 - y^2}} &
\;\;\;\;\;\;\;\; u(u+2)(1-2u/\kappa) & dx\;dy ,\\
d\sigma_\perp & = & - \frac{ \xi_\circlearrowright \zeta_\perp r_e^2}{\kappa(1+u)^3\sqrt{1 - x^2 - y^2}} &
\;\;\;\;\;\;\;\; u  y & dx\;dy .
\end{aligned}
\label{xyall}
\end{equation}
Due to the presence of $\sqrt{1 - x^2 - y^2}$ in the denominator of (\ref{xyall}) the cross section has singularity a the edge of a circle (ellipse),
which however is integrable.

\subsection{The measurements}

The detectors for scattered photons and electrons are going to be installed as it was shown in Fig~\ref{fig:100m}.
These pixel detectors will measure the $x$ and $y$ positions of each particle according to the following scheme:
\begin{figure}[ht]
\centering
\vspace{-4mm}
\includegraphics[width=0.8\textwidth]{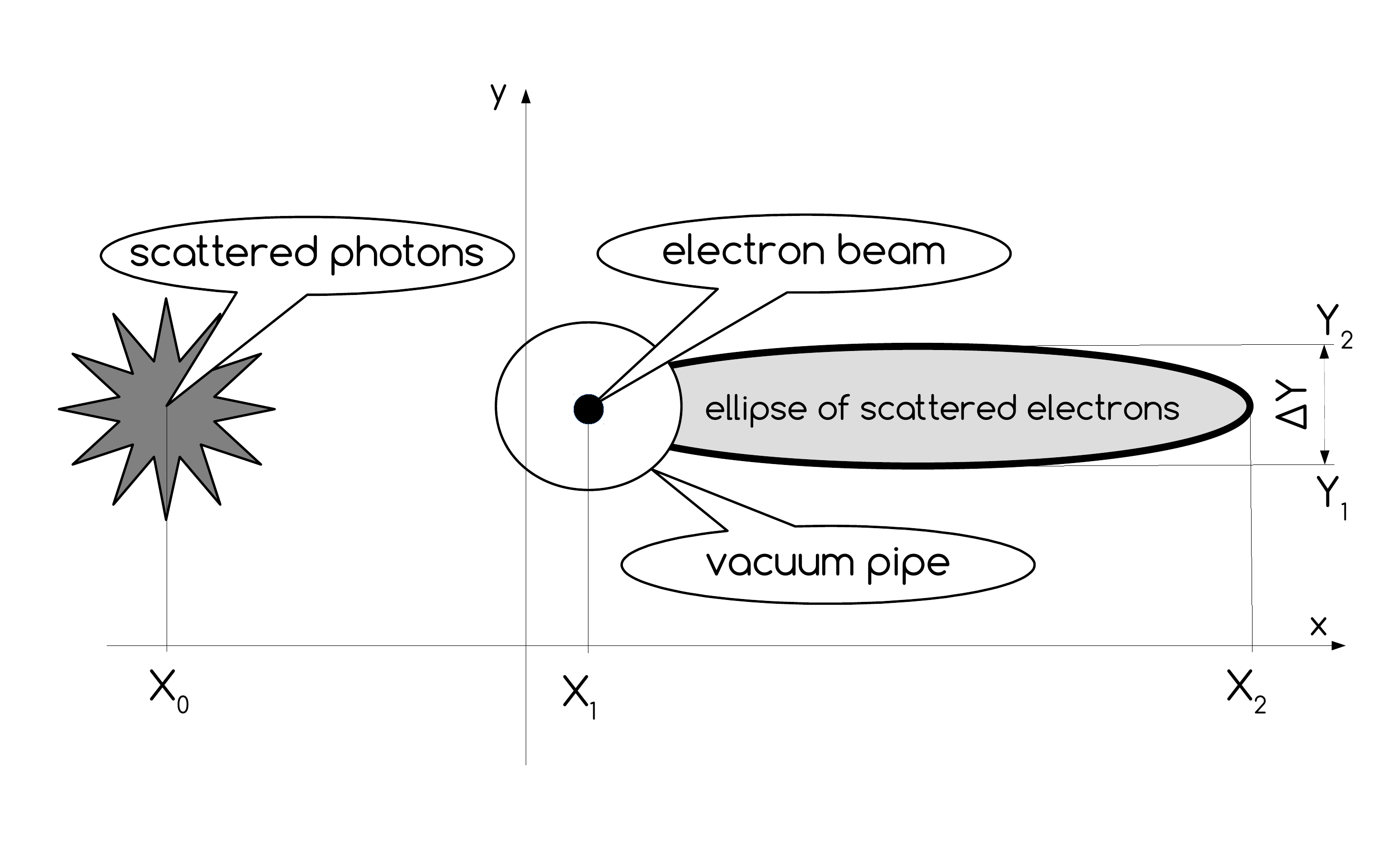}
\vspace{-4mm}
\caption{The $xy$ plane of particle detection. $X_0$ is the horizontal center of gravity in the scattered photons distribution. $X_1$ is the electron beam position and at the same time -- the left edge of the scattered electrons ellipse, while $X_2$ is the right side of the ellipse. Vertical size $\Delta Y = L_1\cdot(4\omega_0/mc^2)$.}
\label{fig:principle}
\end{figure}

For the detection of the scattered electrons we consider a position measurement using a silicon pixel detector (as in \cite{Mordechai:2013zwm}) placed at a distance $L_1= 117$~m from the Compton IP and $L_2 = 100$~m from the center of bending dipole.
The active dimension of the detector is 400$\times$4~mm$^2$, it is shifted horizontally 40~mm away from the beam axis.
The size of the pixel cell taken is 2$\times$0.05~mm$^2$, i.~e. there are 200 pixels in $x$ and 80 pixels in $y$.

We will fit the MC distribution of scattered electrons by theoretical cross section (\ref{xyall}).
This cross section has a very sharp edge at $x^2+y^2=1$, so the integrals of (\ref{xyall}) over each pixel are required for fitting.
However, the dependences of the cross section on $u$ and $y$ are rather weak, so it was found to be enough to take the integral
\begin{equation}
I_{xy} = \int\limits_{x_0}^{x_1}\int\limits_{y_0}^{y_1}\frac{dx\,dy}{\sqrt{1-x^2-y^2}}
\label{Ixy}
\end{equation}
over a rectangular pixel limited by $[x_0,x_1]$ in $x$ and $[y_0,y_1]$ in $y$.

The result of integration is:
\begin{equation}
\begin{aligned}
I_{xy} & =
x_1 \arctan\!\left[\frac{y_1 D_{10}     - y_0 D_{11}}  {y_1 y_0       + D_{11}D_{10}}\right] -
    \arctan\!\left[\frac{x_1(y_1 D_{10} - y_0 D_{11})} {x_1^2 y_0 y_1 + D_{11}D_{10}}\right] +
y_1 \arctan\!\left[\frac{x_1 D_{01}     - x_0 D_{11}}  {x_1 x_0       + D_{11}D_{01}}\right] -  \\
& -
x_0 \arctan\!\left[\frac{y_1 D_{00}     - y_0 D_{01}}  {y_1 y_0       + D_{00}D_{01}}\right] +
    \arctan\!\left[\frac{x_0(y_1 D_{00} - y_0 D_{01})} {x_0^2 y_0 y_1 + D_{00}D_{01}}\right] -
y_0 \arctan\!\left[\frac{x_1 D_{00}     - x_0 D_{10}}  {x_1 x_0       + D_{00}D_{10}}\right],
\end{aligned}
\label{Irect1}
\end{equation}
where $D_{ij}=\sqrt{1-x_i^2-y_j^2}$ and $i,j=[0,1]$.
The second step is to calculate the convolution of $I_{xy}$ with the two-dimensional normal distribution of initial electrons:
$P(x,y) = \frac{1}{2\pi\sigma_x\sigma_y}\exp\left(-\frac{x^2}{2\sigma^2_x}-\frac{y^2}{2\sigma^2_y}\right).$
It is not hard to show, that $\sigma_x$ and $\sigma_y$ are the RMS electron beam sizes (due to betatron and synchrotron motion) at the plane of detection.
And the last step is to account for $u$ and $y$ in eq.~(\ref{xyall}).

\begin{figure}[h!]
\centering
\includegraphics[width=\textwidth]{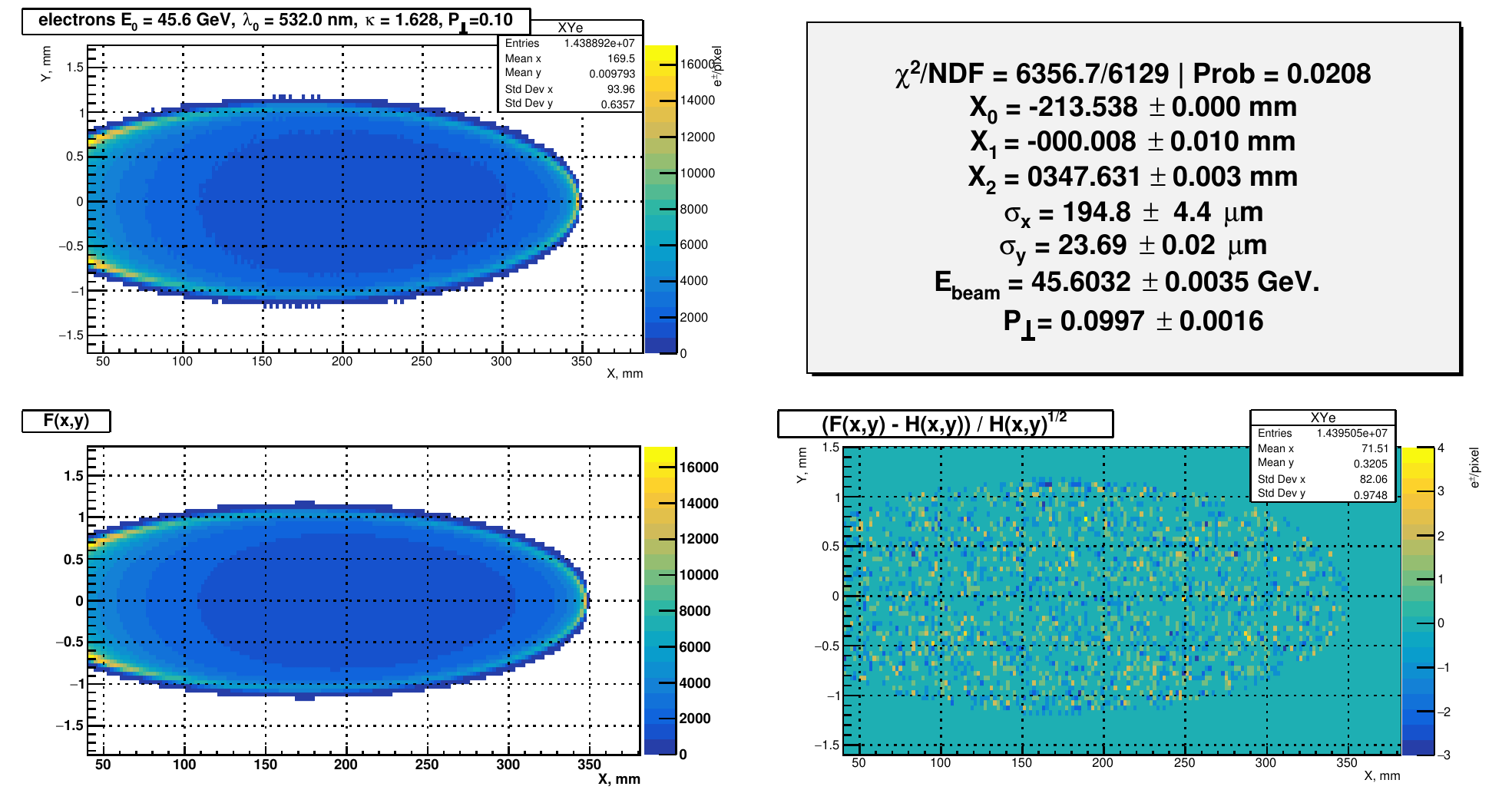}
\caption{{\em Top-left:} MC distribution of scattered electrons $H(x,y)$.
{\em Bottom-left:} function $F(x,y)$ after fitting.
{\em Bottom-right:} normalized difference: $(F(x,y)-H(x,y))/\sqrt{H(x,y)}$.
{\em Top-right:} the $F(x,y)$ parameters obtained by fitting (except $X_0$, which is the mean $x$ value of the scattered photons distribution).}
\label{fig:TheResult}
\end{figure}

The $F(x,y)$ function was built based on these considerations in order to describe the shape of the scattered electrons distribution, see Fig.~\ref{fig:TheResult}.
It has nine parameters (except normalization):
\begin{itemize}
\item The first parameter is $\kappa$, defined in eq.~(\ref{kappa}). This parameter is fixed according to approximate value of the beam energy cause $F(x,y)$ weakly depends on $\kappa$, $\pm$1\% changes does not matter on the fit results.
\item The next four parameters are $X_1,X_2,Y_1,Y_2$ -- positions of the ellipse edges, see Fig.~\ref{fig:principle}.
\item The sixth and seventh are responsible for polarization sensitive terms $P_\perp = \xi_\circlearrowright \zeta_\perp$ and $P_\parallel = \xi_\circlearrowright \zeta_\circlearrowright$. In the example below the fixed conditions are $\phi_\perp = \pi/2$ and $\zeta_\circlearrowright=0$.
\item The eighth and ninth are $\sigma_x$ and $\sigma_y$ -- the electron beam sizes at the azimuth of the detector.
\end{itemize}

The results presented in Fig.~\ref{fig:TheResult} were obtained with $2 \cdot 10^7$ backscattered MC events and about $1.44 \cdot 10^7$ of them was accepted by scattered electrons detector.
The physical parameters obtained directly from the fit results are the ellipse positions $X_1,X_2,Y_1,Y_2$, beam transverse sizes $\sigma_x$ and $\sigma_y$ and the beam polarization degree $P_\perp$, measured with 1.6\% relative accuracy (0.16\% absolute accuracy).
The beam energy is evaluated as:
\begin{equation}
E_{beam} = \frac{(mc^2)^2}{4\omega_0} \cdot \frac{X_2-X_1}{X_1-X_0}.
\end{equation}

\section{The flux of backscattered photons}

Consider CW TEM$_{00}$ laser radiation propagating along $z$-axis.
By definition, the optical radius $w(z)$ of the gaussian beam is the transverse distance from $z$-axis where the radiation intensity drops to $1/e^2$ ($\simeq$13.5\%) from the maximum value.
Let's define the beam size as $\sigma(z) = w(z)/2$ where intensity drops to $1/e$ ($\simeq$36.8\%).
If a laser light of wavelength $\lambda_0$ is focused at $z=0$ to the waist size of $\sigma_0$, the beam size will evolve along $z$:
\begin{equation}
\sigma(z) = \sigma_0 \sqrt{ 1 + \left(\frac {z} {z_R}\right)^2},
\text{ where }
z_R = \frac{4 \pi \sigma_0^2}{\lambda_0} \text{ is the Rayleigh length}.
\end{equation}
The optical intensity [W/cm$^2$] in a Gaussian beam of power $P$ [W] is:
\begin{equation}
I(r,z) = \frac{P}{2 \pi \sigma(z)^2} \exp \left( - \frac {r^2} {2\sigma(z)^2}\right).
\end{equation}
The Rayleigh length is a distance along $z$ from the beam waist where on-axis intensity decreases twice: $I(0,z_R) = I(0,0)/2$.
Far field divergence is $\theta = \sigma_0/z_R = \lambda_0/(4\pi\sigma_0)$.
Radiation power is defined the number of laser photons emitted per second:
\begin{equation}
P = dE/dt = h \nu \cdot dN/dt\;\;[\text{J s}^{-1}].
\end{equation}
Thus the longitudinal density of laser photons along $z$ is:
$\rho_\parallel = dN/dz = P \lambda_0/hc^2$~[cm$^{-1}$].
Consider an electron ($v/c\simeq1$) propagating towards the laser head sea with small incident angle $\alpha$.
\begin{figure}[h!]
\centering
\includegraphics[width=0.5\textwidth]{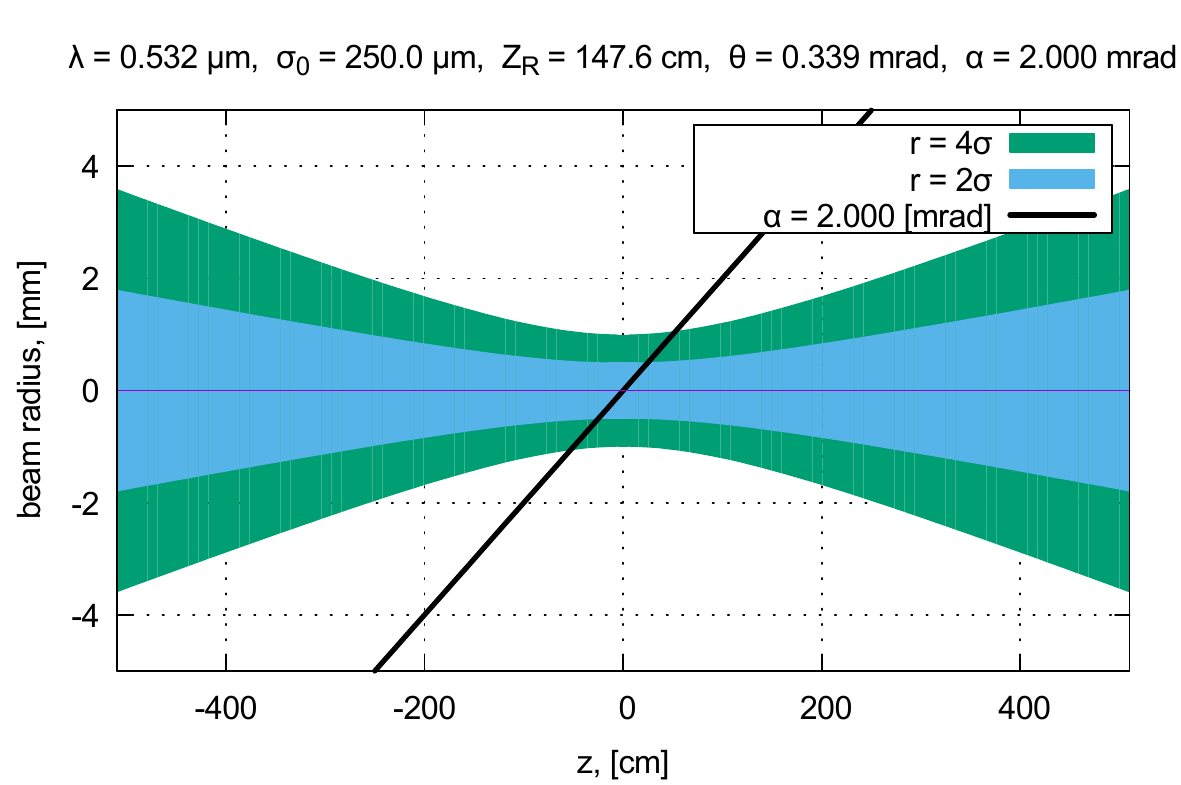}
\caption{An electron (the black sloping line) passing through the laser beam waist.}
\label{fig:beam}
\end{figure}

\begin{figure}[h]
\centering
\includegraphics[width=0.5\textwidth]{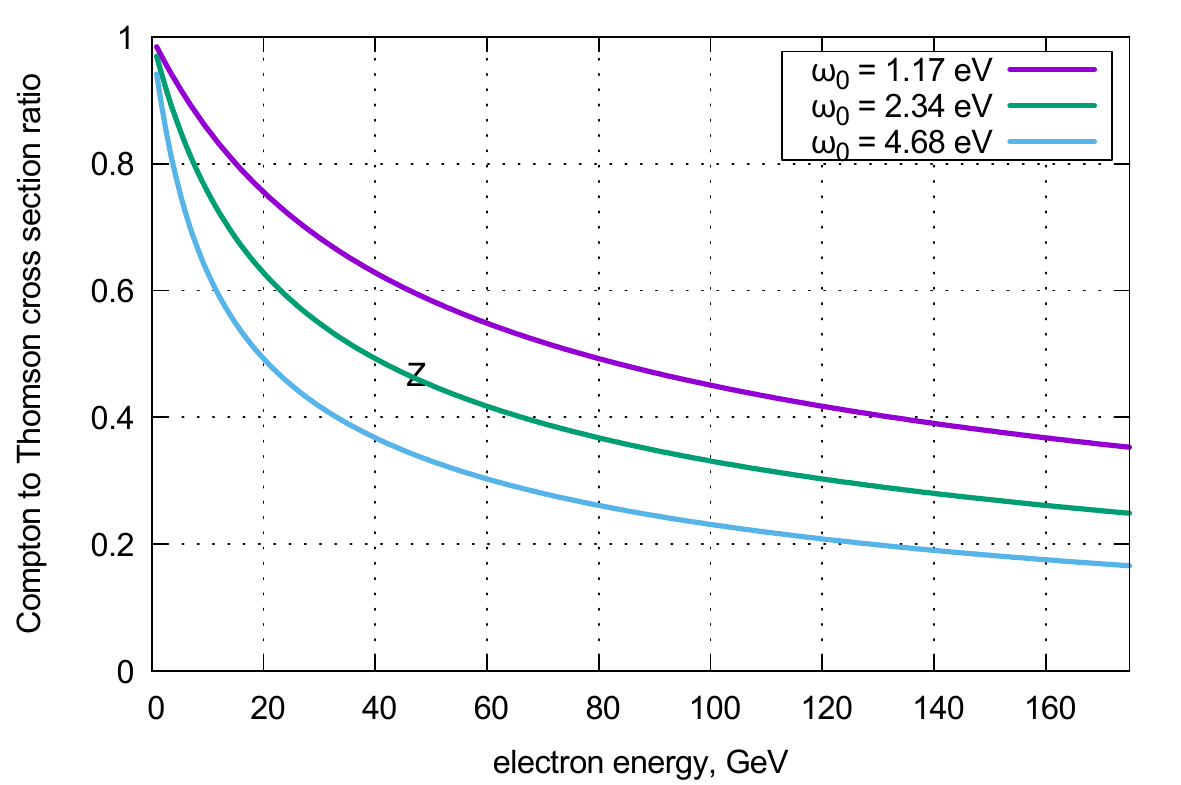}
\caption{The ratio of the ICS cross section to Thomson cross section vs FCC beam energy.}
\label{fig:tcs}
\end{figure}

The photon target density seen by this electron is defined as:
\begin{equation}
\rho_\perp =  \rho_\parallel\frac{(1+\cos\alpha)}{2\pi\sigma_0^2} \int \limits_{-\infty}^{\infty}
\frac{\exp \left( - \frac {z^2 \tan^2\alpha} {2\sigma(z)^2}\right)}{1+(z/z_R)^2} dz\;\;[\text{cm}^{-2}].
\end{equation}
The probability $W$ of the Compton scattering is determined by the product of $\rho_\perp$ and the scattering cross section.
The latter is defined by eq.~(\ref{tcs}) and depends on $\kappa$ parameter, see Fig.~\ref{fig:tcs}.
The maximum scattering probability $W_{max}$ is reached in case $\alpha=0$ and at low energy with Thomson cross section $\sigma_T=0.665$~barn.
\begin{equation}
W_{max} =  \frac{\sigma_T}{\pi\sigma_0^2}\frac{P \lambda_0}{ h c^2 } \int \limits_{-\infty}^{\infty}\frac{dz}{ 1 + (z/z_R)^2}
 = \frac{4 \sigma_T P}{ h c^2 } \int \limits_{-\infty}^{\infty}\frac{dx}{ 1 + x^2} = \frac{4 \pi \sigma_T P}{ h c^2 } = \frac{P}{P_c},
\end{equation}
where $ P_c = \hbar c^2/2\sigma_T \simeq 0.7124 \cdot 10^{11}$~[W] is the power of laser radiation required for 100\% scattering probability.
We see that $W_{max}$ depends neither on the radiation wavelength $\lambda_0$ nor the waist size $\sigma_0$, but on the laser power only.
A low energy electron bunch with $0.7\cdot10^{11}$ population colliding head-on with 1~W of laser radiation will produce one Compton scattering event.

The loss in scattering probability when $\alpha \neq 0$ is defined by the ratio of angle $\alpha$ to the laser divergence angle $\theta=\sigma_0/z_R$.
Since the mirror is required in order to deliver the laser beam to IP, $\theta$ should be always smaller than $\alpha$: this ratio finely will describe the laser and electron beam separation at the place of mirror installation (see Fig.~\ref{fig:100m}).
If we define the ``Ratio of Angles'' as $R_{A} = \alpha/\theta$, probability loss will be expressed as:
\begin{equation}
\eta(R_A) = \frac{W(\alpha)}{W_{max}} = \frac{1}{\pi} \!\!\int \limits_{-\infty}^{\infty} \!\!\exp \left( - \frac{x^2 R_A^2}{2(1+x^2)}  \right) \frac{dx}{ 1 + x^2}.
\label{etara}
\end{equation}
\begin{figure}[h]
\centering
\vspace{-0.5cm}
\includegraphics[width=0.45\textwidth]{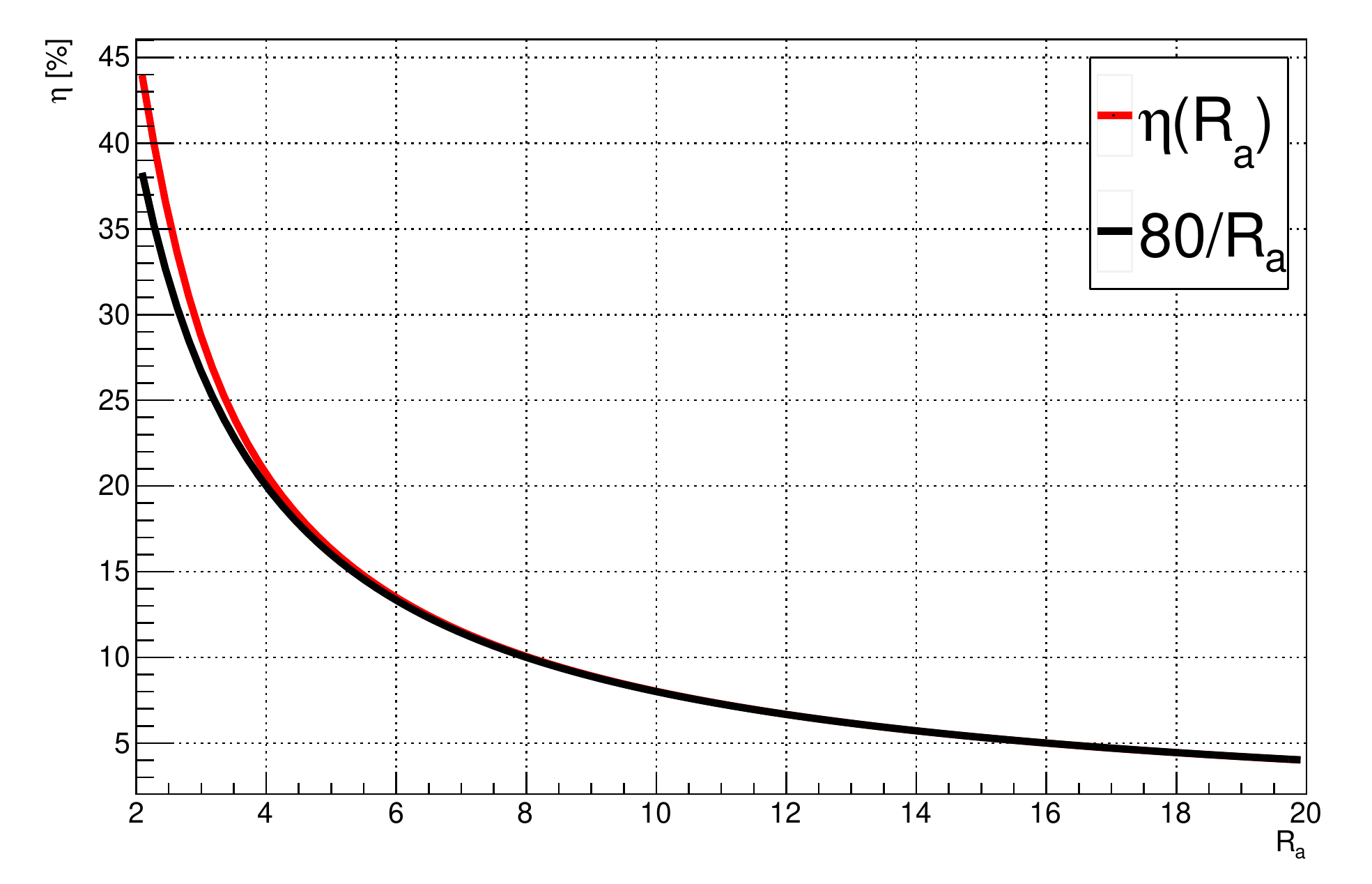}
\caption{$\eta(R_A)$ vs $R_A$ as defined by eq.~(\ref{etara}) and its simple approximation.}
\label{fig:eta1}
\end{figure}

\subsection{Pulsed laser}

At the FCC-ee there will be polarized pilot bunches for regular beam energy measurement by resonant depolarization.
So the laser system should provide the backscattering on a certain electron bunch, and laser operation in CW mode is thus not possible.
The FCC-ee revolution frequency $\simeq 3$~kHz is comfortable for solid-state lasers operating in a Q-switched regime.
The laser pulse propagation can be described as:
\begin{equation}
\rho_\parallel(s,t) = \displaystyle\frac{N_\gamma}{\sqrt{2\pi}c\tau_L}\exp\left(-\frac{1}{2}\left(\frac{s-ct}{c\tau_L}\right)^2\right),
\end{equation}
where $\tau_L$ and $E_L$ are pulse duration and energy, $N_\gamma = E_L\lambda/hc$.
Scattering probability for $\alpha=0$ is:
\begin{equation}
W  =  \displaystyle\frac{(E_L/\sqrt{2\pi}\tau_L)}{P_c} \cdot
\frac{1}{\pi}\int\limits_{-\infty}^{\infty}\!\frac{\exp\{-2(x\,z_R/c\tau_L)^2\}}{1+x^2}
\,dx =
\displaystyle\frac{P_L}{P_c} \cdot \frac{1}{\pi}\int\limits_{-\infty}^{\infty}\frac{\exp\{-2(x R_L)^2\}}{1+x^2}dx,
\end{equation}
where $P_L = E_L/\sqrt{2\pi}\tau_L$ is the instantaneous laser power and $R_L=z_R/c\tau_L$ is the ``Ratio of Lengths''.

The scattering probability for an arbitrary $\alpha$ is:
\begin{equation}
W  =  \displaystyle\frac{P_L}{P_c} \cdot \frac{1}{\pi}
\int\limits_{-\infty}^{\infty}\frac{\exp\left(-x^2\left(2 R_L^2+\displaystyle\frac{R_A^2}{2(1+x^2)}\right)\right)}{1+x^2}\,dx =
\displaystyle\frac{P_L}{P_c} \cdot \eta(R_L, R_A)
\label{nlna}
\end{equation}
where
$$
P_L = E_L/\sqrt{2\pi}\tau_L; \;\;\;\;\;\;
P_c \simeq 0.7124 \cdot 10^{11}~\text{[W]}; \;\;\;\;\;\;
R_L=z_R/c\tau_L; \;\;\;\;\;\;
R_A = \alpha/\theta_0 = 4\pi\sigma_0/\alpha.
$$

The map of the efficiency $\eta(R_L, R_A)$, obtained by numerical integration of eq.~(\ref{nlna}), is presented in Fig.~\ref{fig:nlna}:
\begin{figure}[h]
\centering
\includegraphics[width=0.75\textwidth]{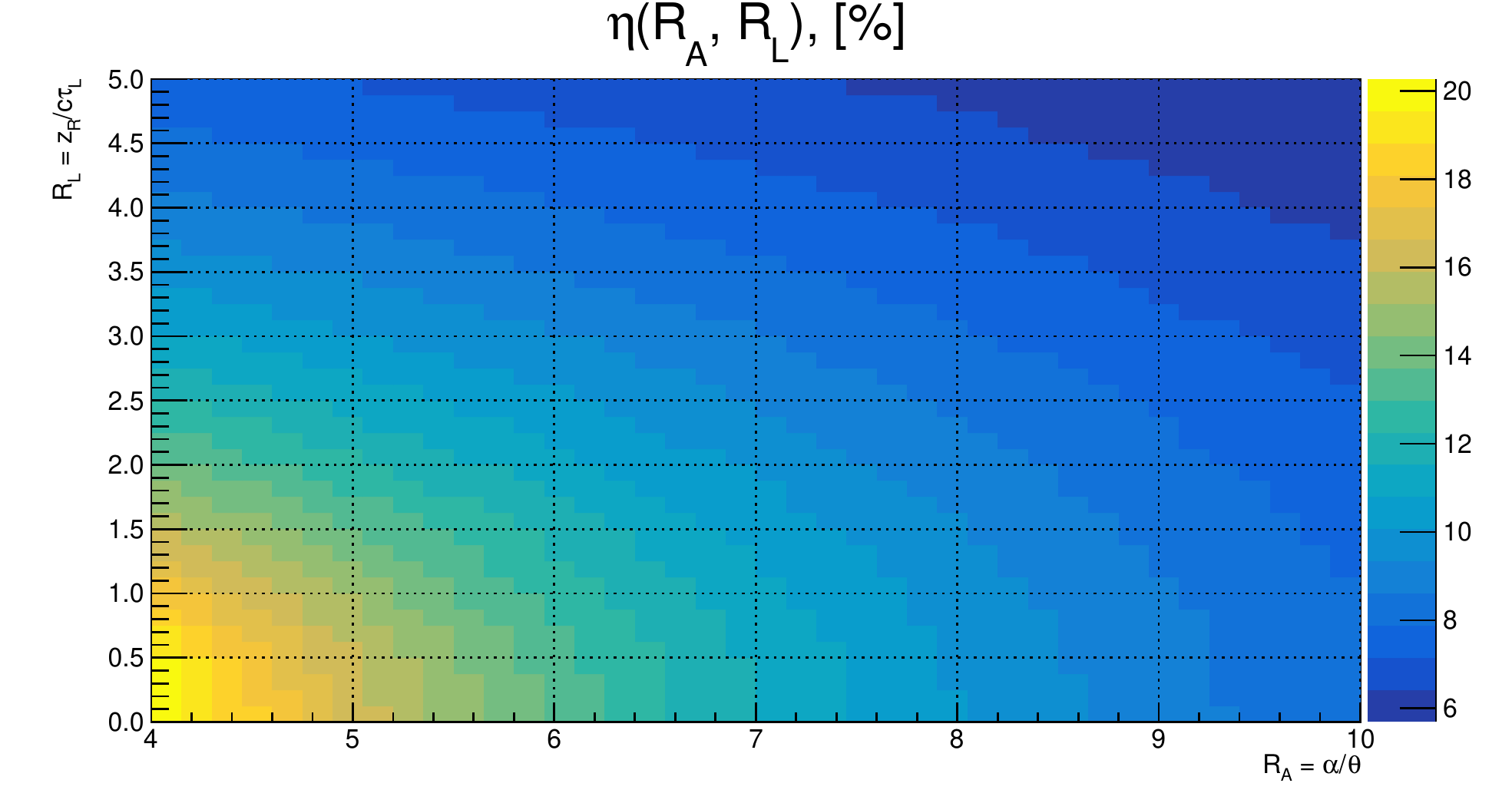}
\label{fig:nlna}
\end{figure}

Now we have enough instruments to estimate the flux of backscattered photons, obtained from one FCC-ee bunch in the configuration, shown in Fig.~\ref{fig:100m}.

\begin{itemize}
\item Laser wavelength $\lambda_0$ = 532~nm.
\item Compton cross section correction (see letter $Z$ on Fig.~\ref{fig:tcs}): $R_\times\simeq$50\%.
\item Waist size $\sigma_0$ = 0.25~mm, Rayleigh length $z_R$ = 148~cm.
\item Far field divergence $\theta$ = 0.169~mrad.
\item Interaction angle $\alpha$ = 1.0~mrad (horizontal crossing).
\item Laser pulse energy: $E_L$ = 1 [mJ], pulse length: $\tau_L$=5~[ns] (sigma).
\item Instantaneous laser power:  $P_L$ = 80 [kW], $P_L/P_c = 1.1 \cdot 10^{-6}$.
\item Ratio of angles $R_A$ = 5.9, ratio of lengths $R_L$ = 0.98.
\item $\eta(R_L, R_A)$ $\simeq$13\% (see Fig.~\ref{fig:nlna}).
\item Scattering probability $W = P_L/P_c \cdot R_\times \cdot \eta(R_L, R_A) \simeq 7 \cdot 10^{-8}$.
\item With $N_e = 10^{10}$ electrons/bunch and $f=3$~kHz repetition rate: $\dot{N}_\gamma = f \cdot N_e \cdot W\simeq 2\cdot 10^{6}$ [s$^{-1}$].
\item Average laser power is $P=f \cdot E_L\simeq3$~W.
\end{itemize}
The influence of the electron beam sizes on the above estimations was not considered cause it is negligible.

\section{Summary}
The electron beam polarimeter for the FCC-ee project has been considered.
With the laser system parameters, described in the latter section, it allows to measure transverse beam polarization with required 1\% accuracy every second.
With the suggested scheme, this apparatus can also measure the beam energy, longitudinal beam polarization, beam position and transverse beam sizes at the place of installation.
The statistical accuracy of direct beam energy determination is $\Delta E/E < 100$~ppm within 10s measurement time.
However, the possible sources of systematical errors require additional studies.
The best case of such studies would be the experimental test of the suggested approach on low-emittance and low-energy electron beam.

\bibliographystyle{unsrt}
\bibliography{polarimeter}
\end{document}